\documentclass[twocolumn]{aastex6}

\begin{document}

\title{Discovery of a Makemakean Moon}

\author{Alex H. Parker\altaffilmark{1} and Marc W. Buie}
\affil{Southwest Research Institute \\
1050 Walnut St., Suite 300 \\
Boulder, CO 80302, USA}

\author{Will M. Grundy}
\affil{Lowell Observatory, Flagstaff, AZ, USA.}

\author{Keith S. Noll}
\affil{NASA Goddard Space Flight Center, Greenbelt, MD, USA.}

\altaffiltext{1}{aparker@boulder.swri.edu}

\begin{abstract}
We describe the discovery of a satellite in orbit about the dwarf planet (136472) Makemake. This satellite, provisionally designated S/2015 (136472) 1, was detected in imaging data collected with the Hubble Space Telescope's Wide Field Camera 3 on UTC April 27, 2015 at 7.80$\pm$0.04 magnitudes fainter than Makemake. It likely evaded detection in previous satellite searches due to a nearly edge-on orbital configuration, placing it deep within the glare of Makemake during a substantial fraction of its orbital period. This configuration would place Makemake and its satellite near a mutual event season. Insufficient orbital motion was detected to make a detailed characterization of its orbital properties, prohibiting a measurement of the system mass with the discovery data alone. Preliminary analysis indicates that if the orbit is circular, its orbital period must be longer than 12.4 days, and must have a semi-major axis $\gtrsim$21,000 km. We find that the properties of Makemake's moon suggest that the majority of the dark material detected in the system by thermal observations may not reside on the surface of Makemake, but may instead be attributable to S/2015 (136472) 1 having a uniform dark surface. This ``dark moon hypothesis'' can be directly tested with future JWST observations. We discuss the implications of this discovery for the spin state, figure, and thermal properties of Makemake and the apparent ubiquity of trans-Neptunian dwarf planet satellites.
\end{abstract}


\section{Introduction} \label{sec:intro}

Makemake is the second-brightest known trans-Neptunian Object (behind only Pluto) and the largest known classical Kuiper Belt Object (KBO). It has the most methane-dominated spectrum of any known TNO \citep{lic06, bro07, teg08}, a very high visible albedo \citep{lim10}, a well-defined radius derived from stellar occultations \citep{ort12, bro13}, a well-measured (albeit very small-amplitude) lightcurve that pins its rotational period to 7.771 hours \citep{hei09}, and polarization properties very similar to Pluto and Eris, but distinct from smaller KBOs \citep{bel12}. However, despite this wealth of information, the lack of a known satellite has prohibited the measurement of Makemake's mass and density.

\begin{figure*}
\centering
\figurenum{1}
    \includegraphics[width=1\textwidth]{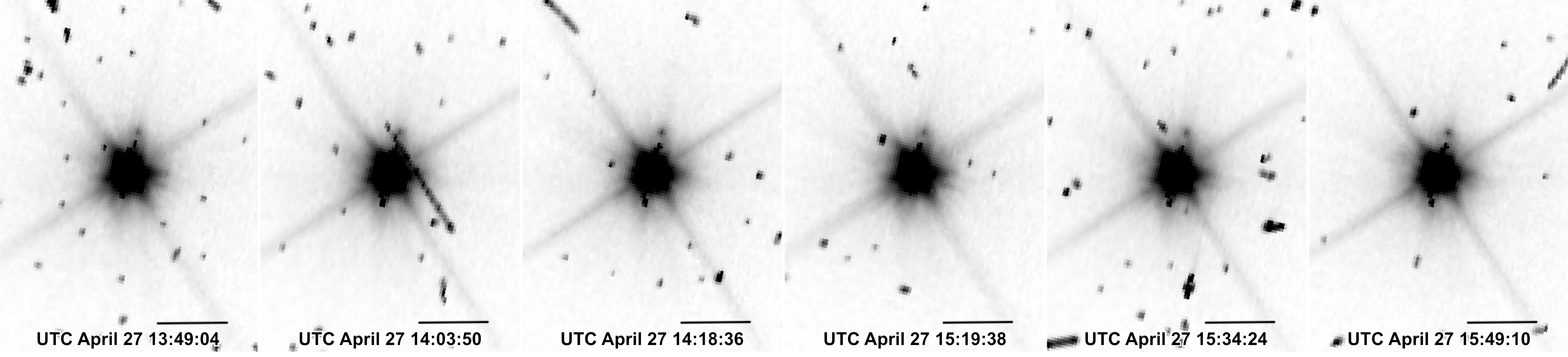}
\caption{All six 725-second images from visit 1, collected over two subsequent HST orbits. The satellite is clearly visible to the North and West of Makemake in every frame. Images have not been cleaned for cosmic rays or otherwise cosmetically enhanced. Black bar is 1\arcsec. Ecliptic North is up, Ecliptic East is left. Observation start times are labeled on each image. Over the time spanned by these images, Makemake moved 4\arcsec.22 with respect to background sources. \label{fig1}}
\end{figure*}

We report the discovery and preliminary characterization of a Makemakean moon in Hubble Space Telescope (HST) Wide-Field Camera 3 (WFC3) imagery. While the satellite's orbital properties are only marginally constrained from the discovery data alone, its existence will permit future precise measurement of Makemake's mass and density given sufficient follow-up observations. In the following sections, we describe the discovery circumstances of S/2015 (136472) 1, its photometric properties, and preliminary characterization of its orbital properties. We demonstrate that this moon could account for some or all of the dark material detected in the Makemake system with thermal observations, and discuss potential future avenues of research enabled by this satellite. We argue that Makemake is likely currently viewed equator-on and derive implications for its figure and thermal properties. We conclude with a discussion of the properties of dwarf planet satellites.

\section{Discovery Circumstances} \label{sec:dc}

As part of the HST GO program 13668, Makemake was imaged with WFC3 in the broad F350LP filter over the course of two visits of two back-to-back orbits each. These visits were separated by approximately two days. Each visit was bracketed by single 12-second images in which Makemake does not saturate; the remainder of each visit was filled with six 725 second exposures in which Makemake saturates. These observations were designed to enable the detection of satellites fainter than could have been found in previous satellite search programs. 

\begin{figure*}[t]
\centering 
\figurenum{2}
    \includegraphics[width=0.75\textwidth]{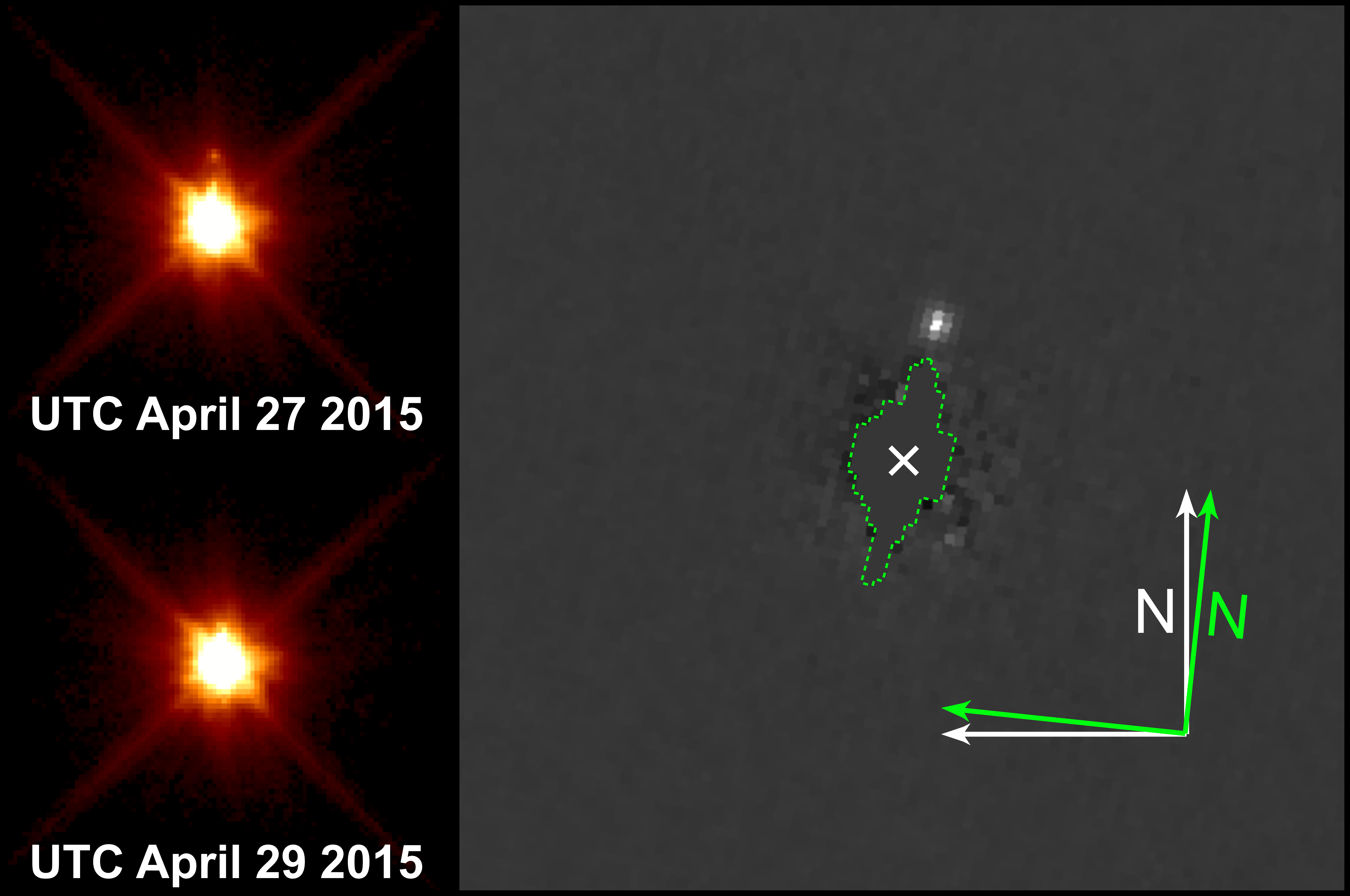}
\caption{Left panels: Co-registered stack of all visit 1 images (top) and visit 2 images (bottom). Images are displayed in their original array coordinates with identical stretch to best compare PSF structure. Right: WCS-rotated stack of visit 1 images with co-registered visit 2 images subtracted, showing S/2015 (136472) 1. Stack is 33rd percentile of six input images. Arrows indicate 1\arcsec in Ecliptic North and East for each visit; white for visit 1 and green for visit 2. Green trace indicates masked region where S/2015 (136472) 1 would not have been recoverable in visit 2 as determined by injecting synthetic sources. White cross indicates centroid of Makemake. \label{fig2}}
\end{figure*}

Visit 1 was on April 27, 2015, from UTC 13:46:36 to 16:03:58, and Visit 2 on April 29, 2015, from UTC 18:17:46 to 20:35:03. In all six 725 second images collected in Visit 1, a faint source is visible 0\arcsec.57 from Makemake; see Figure \ref{fig1}. Over the 132-minute duration spanned by these observations, Makemake moved 4\arcsec.22 with respect to background sources (more than 60 times the F350LP point-spread function FWHM); the fainter source was precisely co-moving with Makemake over this period. The source was not visible in Visit 2 (see Figure \ref{fig2}), and subsequent efforts to reduce the confusion produced by Makemake's PSF through difference imaging did not reveal the source. 

We injected synthetic PSFs into the Visit 2 difference images in order to determine our sensitivity to a source and the implications of a non-detection. We find that sources up to 2.5 magnitudes fainter than the Visit 1 source are reliably recovered at large separations from Makemake, and sources one magnitude fainter are visible in all regions but the saturated core of Makemake (illustrated by the masked region in Figure \ref{fig2}). Previous HST observations of Makemake taken with the Advanced Camera for Surveys (ACS) High Resolution Channel on November 19, 2006, were of sufficient sensitivity to detect this satellite (program GO 10860 included 16 550 second exposures in F606W; 5$\sigma$ limiting magnitude in each from estimated from the ACS exposure time calculator is $V\sim26$); however, the satellite is not visible in these data either. We infer from these two non-detections that the satellite spends a large fraction of its time very close to Makemake in the sky plane, likely in an edge-on orbit. We discuss this further in Section \ref{sec:astrom}.

\subsection{Photometry} \label{sec:phot}

For each of the six frames in which S/2015 (136472) 1 is visible, we subtracted a co-registered median stack of the six frames in which it was not visible. In five of these six difference images, we performed small-aperture photometry to measure the flux from S/2015 (136472) 1; in the sixth, a cosmic ray impinged too close to S/2015 (136472) 1. The typical signal-to-noise of S/2015 (136472) 1 in these frames is $S/N\sim25$. The results are listed in Table 1. Four bracketing short F350LP exposures were also median stacked (without differencing) to measure the flux from Makemake itself in the filter passband. We find that S/2015 (136472) 1 is 7.80$\pm$0.04 magnitudes fainter than Makemake in F350LP. In the absence of any color information on the satellite, we adopt this delta-magnitude for V-band photometry. Makemake is $H_{v}=0.091\pm0.015$ \citep{rab07}, and from this we estimate $H_v=7.89\pm0.04$ for S/2015 (136472) 1. At the time of discovery, the system was at a heliocentric range of 52.404 AU, a geocentric range of 51.694 AU, and observed at a phase angle of 0.781$^\circ$.

\subsection{Astrometry and Orbital Properties} \label{sec:astrom}

We performed preliminary orbit modeling to determine the possible range of system parameters given the limited astrometric information derived from the discovery data. To avoid over-fitting the data, we assume a prograde circular orbit for preliminary estimates and adopt a simple pass-fail criterion: if, given a set of orbit parameters, the predicted positions of S/2015 (136472) 1 at the times with measured astrometry (Table \ref{tab1}) are within 0\arcsec.016 (twice the estimated raw astrometric uncertainty of 0\arcsec.008, given $S/N\sim25$, the dithering precision of HST, and the undersampled WFC3 PSF) from the measured locations at those times, and the predicted positions of S/2015 (136472) 1 at the times when it was \emph{not} detected fall within the masked region in Figure \ref{fig2}, then an orbit is accepted as plausible given the discovery data alone. We densely sampled a large volume of orbital parameter space and determined the maximum plausible range of each orbital parameter of interest under the stated assumptions above.

We sampled bulk density for Makemake over the range 1.4 g cm$^{-3}$ $\leq \rho \leq$ 3.2 g cm$^{-3}$ \citep{bro13}, and find that the S/2015 (136472) 1 discovery observations alone do not further constrain Makemake's density. Given this range of densities, semi-major axes in 21,100 km $\lesssim$ a $\lesssim$ 300,000 km, inertial orbital periods in 12.4 days $\lesssim \tau \lesssim$ 660 days, and prograde Ecliptic inclinations in $63^{\circ} \lesssim i_E \lesssim 87^{\circ}$ are acceptable (the mutual inclination of the system is $46^{\circ} \lesssim i_M \lesssim 78^{\circ}$, and the inclination with respect to the sky plane is $83^{\circ} \lesssim i_S \lesssim 105^{\circ}$; the retrograde mirror solutions to all of these ranges is also acceptable). 

Given the existence of previous HST satellite search data of sufficient depth to detect S/2015 (136472) 1, we consider the largest semi-major axis solutions unlikely due to the fact that these orbital configurations place S/2015 (136472) 1 at large separations from Makemake for the majority of the time; with a=100,000 km, S/2015 (136472) 1 spends $\sim$90\% of the time at detectable separations, whereas with the minimum allowable circular semi-major axis of 21,000 km, S/2015 (136472) 1 spends only $\sim$50\% of the time at detectable separations. On the other hand, semi-major axes in excess of 100,000 km are known to exist for much less massive binary Kuiper Belt Objects (eg., 2001 QW322, \citealt{par11}). The largest semi-major axis solutions are found for the highest adopted Makemake density, and these solutions are at $\sim$50\% of the Makemake Hill radius for this density.

Given the equations in \citet{nol08}, if the orbit of S/2015 (136472) 1 has a semi-major axis near its lower limit and the components both have $\rho=2$ g cm$^-3$, the orbital circularization timescale for a D=175 km satellite is approximately 60 Myr, while if the semi-major axis is twice this lower limit, the orbital circularization timescale increases to longer than the age of the solar system, $\sim6$ Gyr. Thus, if S/2015 (136472) 1 is in an orbit with semi-major axis consistent with its discovery separation, its orbit is very likely circular; if the semi-major axis is much larger, eccentric orbits become more plausible. 

\clearpage

\section{Discussion} \label{sec:disc}

\subsection{Possibility of serendipitous alignment of an unbound TNO}

Given the limited amount of orbital motion about Makemake observed for S/2015 (136472) 1, we must consider the possibility that the detection is a false positive arising from another TNO serendipitously crossing the same line of sight as Makemake. We can strongly rule out this scenario with the following simple analysis.
At the time of the discovery observations, Makemake was at a heliocentric Ecliptic latitude of $\sim28.5^\circ$ and a heliocentric range of $\sim$52 AU. Since the parallax produced over the width of HST's orbit will vary by nearly a full UVIS WFC3 pixel for objects within $\sim$5 AU of 52 AU, we cap the heliocentric range of potential coincident TNOs to 47---57 AU. Using the \emph{Canada France Ecliptic Plane Survey} L7 synthetic model of the Kuiper Belt \citep{pet11}, we find that the typical sky density of Hv $\leq$ 7.8 TNOs with 28$^\circ$ $\leq$ $\beta$ $\leq$ 29$^\circ$ and 47 $\leq$ R $\leq$ 57 is $\sim0.1$ deg$^{-2}$. Given three relatively distinct epochs (counting HST observations from program 10860 collected in November 2006 which did not detect the satellite) of observations capable of detecting S/2015 (136472) 1, the odds of a similarly-bright or brighter TNO at a similar heliocentric distance falling within an arcsecond of Makemake in one or more epoch is less than one in 10$^7$.

\subsection{Dark Moon Hypothesis}

Thermal observations of Makemake collected by the Spitzer \citep{sta08} and Herschel \citep{lim10} space telescopes revealed that there were at least two distinct surfaces contributing to the spectral energy distribution; the majority of the emitting surface must be very bright, but a small component must be very dark. It was argued that the presence of distinct dark terrains on the surface of Makemake was at odds with Makemake's very small lightcurve amplitude unless Makemake is in a pole-on viewing geometry \citep{bro13}. 

Given the discovery of S/2015 (136472) 1, we reconsider these thermal observations under the hypothesis that some or all of the dark material in the system does not reside on the surface of a modestly mottled Makemake, but rather covers the entire surface of a uniformly dark S/2015 (136472) 1. Lim et al. (2010) require a dark surface area with equivalent diameter 310 km $< D <$ 380 km, with geometric albedo $0.02 < p_v < 0.12$. For $p_v=0.02$, the estimated H-magnitude of S/2015 (136472) 1 corresponds to a diameter of $\sim$250 km, too small to account for all of the dark material, but sufficient to account for a large fraction of it. This would reduce the preference for a pole-on orientation for Makemake's rotation. A very dark surface might suggest an origin distinct from other dwarf planet satellites, perhaps indicating that the satellite is a captured, formerly-unbound TNO. Alternatively, a dark satellite surface may be the result of past epochs of interactions with an escaping Makemake atmosphere.

\begin{figure}[t]
\centering
\figurenum{3}
    \includegraphics[width=\columnwidth]{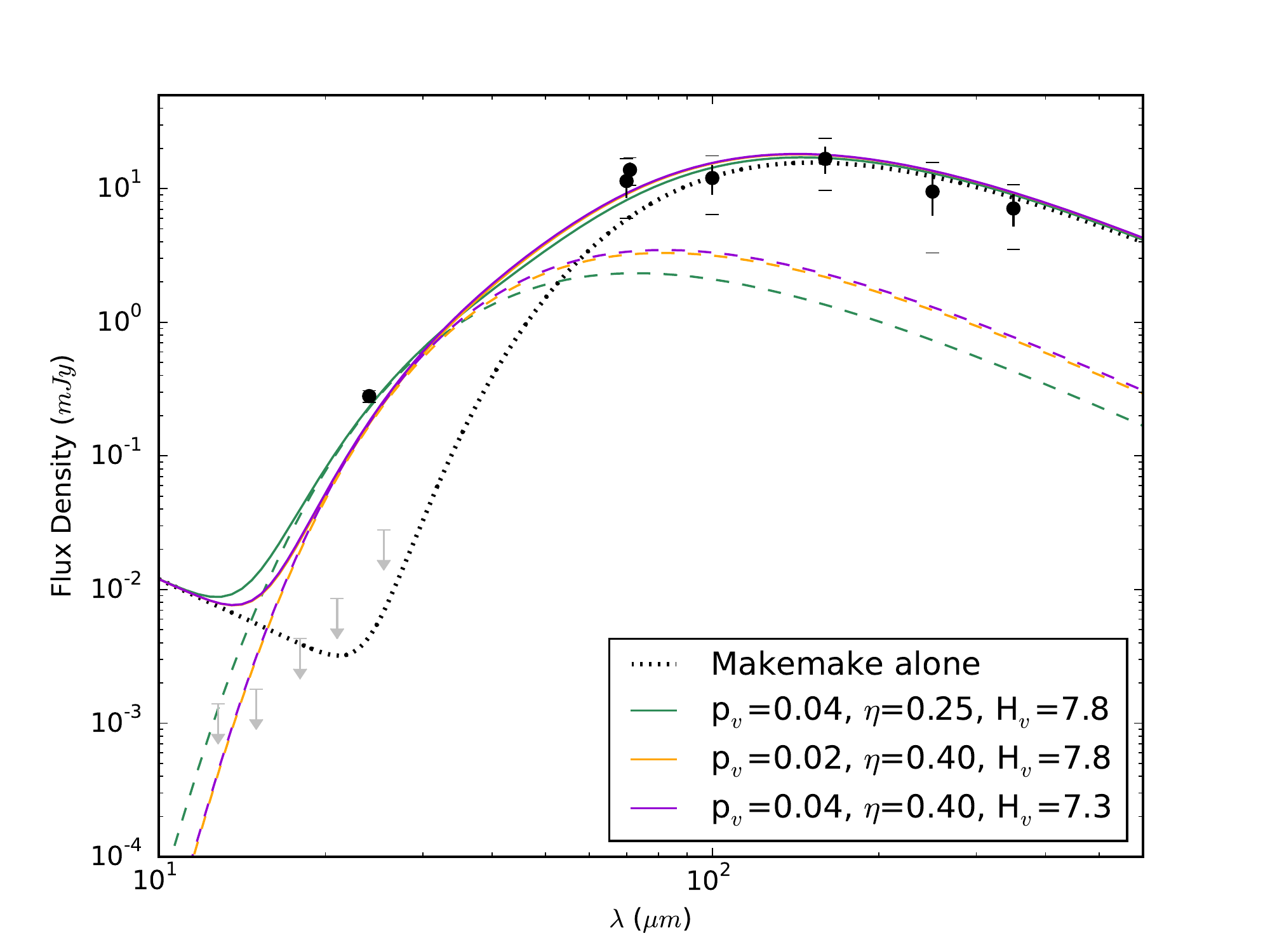}
\caption{Models of thermal emission from Makemake and S/2015 (136472) 1 compared to observations from Herschel (Lim et al. 2010) and Spitzer (Stansberry et al. 2008). Black points: observed data, corrected to common geometry of the later Herschel observations, with one- and two-$\sigma$ error ranges illustrated. Black dotted line: Model Makemake thermal emission, with D=1430km, $\eta=1.9$, and $p_v$=0.82. Dashed colored lines illustrate three models of S/2015 (136472) 1 thermal emission, and solid colored lines illustrate sum of Makemake and S/2015 (136472) 1 model emission. None of the selected models fully reproduce the measured 24$\mu$m flux density measured by Spitzer; either a more extreme surface for S/2015 (136472) 1 is required, or the surface of S/2015 (136472) 1 is not the only dark material in the system. Gray arrows illustrate the JWST MIRI 10,000 second 10$\sigma$ detection limits for 12.8$\mu$m, 15$\mu$m, 18$\mu$m, 21$\mu$m, and 25.5$\mu$m filters.\label{fig3}}
\end{figure}

As a simple check, we use NEATM \citep{har98} to model the thermal flux from the system with the occultation-updated diameter of 1430 km for Makemake and three different surfaces for S/2015 (136472) 1, and compare them against previous thermal observations of the system. Makemake's surface is modeled with a uniform geometric albedo of 0.82, and a beaming parameter of $\eta=1.9$. The three models of S/2015 (136472) 1 include (a) a model with Salacia-like 4\% geometric albedo, but an exceptionally low beaming parameter of $\eta=0.25$, (b) a model with a low 2\% geometric albedo, but a beaming parameter in-family with the results of \citet{lim10}, $\eta=0.4$, and (c) a model with a 4\% geometric albedo and a beaming parameter $\eta=0.4$, but with an absolute magnitude 0.5 magnitudes brighter than observed for S/2015 (136472) 1, capturing the possibility that S/2015 (136472) 1 has a substantial lightcurve. The results are illustrated in Figure \ref{fig3}. These three models all produce between 60\%---80\% of the measured 24$\mu$m spectral flux density from the system.

\subsection{Future Observations}

Follow-up observations of S/2015 (136472) 1 will permit the measurement of Makemake's mass and density. Since we cannot yet predict the future position of S/2015 (136472) 1 with respect to Makemake, it is likely that any recovery efforts will be hampered by an initial period in which S/2015 (136472) 1 is lost in a large fraction of observations before the orbit can be sufficiently well modeled and the recovery rate increased. With sufficient recovery observations, the system mass will be measured. Given a nominal geometric albedo of 4\%, S/2015 (136472) 1 would have a diameter of 175 km and for equal densities would contribute $\lesssim0.2$\% of the system mass; thus, the system mass will be dominated by Makemake.

Because of the nearly edge-on configuration of the orbit, there is the potential for a near-future epoch of mutual events between S/2015 (136472) 1 and Makemake. As with the mutual events between Pluto and Charon in the late 1980s (e.g., \citealt{bui92, you01}), such a configuration could enable detailed investigations of the system and the surface properties of the two components.

Future observations with JWST can test the dark moon hypothesis; \citet{par11} highlight that JWST could characterize the anomalous thermal excess of Makemake in detail. Figure \ref{fig3} illustrates the 10,000 second 10$\sigma$ MIRI detection limits in its five longest-wavelength wide filters. At 15$\mu$m, Makemake and all three S/2015 (136472) 1 model surfaces produce detectable spectral flux density, and at this wavelength the spatial resolution of JWST is comparable to the discovery separation of Makemake and S/2015 (136472) 1. Resolved JWST observations in this and the two adjacent filters would enable direct determination of the fraction of the system's dark material that is attributable to the surface of S/2015 (136472) 1. However, the success of such observations relies upon the ability to predict the sky-plane separation of Makemake and S/2015 (136472) 1, requiring that near-term observations of S/2015 (136472) 1 be conducted to accurately measure its orbital properties.

\subsection{The Figure, Obliquity, and Thermal State of Makemake}

Makemake's high rate of rotation makes its equilibrium figure a Maclaurin spheroid \citep{ort12}. Given even a small flattening, Makemake's J$_2$ would likely drive the system to a low obliquity between the satellite's orbit and the spin pole of Makemake (e.g., \citealt{por12}). We note that the projected long axis of Makemake measured by \citep{ort12} runs nearly North-South, which is consistent with our determination of the orientation of the orbit plane of S/2015 (136472) 1 --- and thus consistent with a low mutual obliquity. If the spin pole of Makemake and the orbit plane of S/2015 (136472) 1 are aligned, we are viewing Makemake nearly equator-on, and the sky-plane elliptical fit presented in \citet{ort12} likely reflects the true axial ratio of Makemake, implying a flattening of $\sim$15\%. If the dark moon hypothesis is correct, then edge-on rotation for Makemake is not at odds with its low-amplitude lightcurve, and a low-amplitude lightcurve in this configuration also implies a rotationally-symmetric, close-to-equilibrium figure.

Additionally, if the spin pole of Makemake and the orbit plane of S/2015 (136472) 1 are aligned, Makemake has a very \emph{high} obliquity with respect to its heliocentric orbit (46$^{\circ}$ --- 78$^{\circ}$). A current edge-on configuration places Makemake near equinox, and if so, we estimate its thermal parameter $\Theta$ \citep{spe90} to be $\sim70$ (given a Pluto-like thermal inertia of 20 J m$^{-2}$s$^{-0.5}$K$^{-1}$, \citealt{lel11}), placing it solidly in the regime in which fast rotator approximations apply. Given this high obliquity, however, as Makemake continues around its orbit, it will effectively transition into a \emph{slow} rotator state at its solstices. Makemake is currently near aphelion, so this potential transition from fast- to slow-rotator is also currently synched with its heliocentric distance extrema. This could lead to fascinating seasonal evolution on Makemake's volatile-dominated surface.

\subsection{The Satellites of Dwarf Planets}

With the discovery of S/2015 (136472) 1, all four of the currently-designated trans-Neptunian dwarf planets (Pluto, Eris, Makemake, and Haumea) are known to host one or more satellites. The fact that Makemake's satellite went unseen despite previous satellite searches suggests that other very large trans-Neptunian objects that have already been subject to satellite searches (such as Sedna and (225088) 2007 OR$_{10}$) may yet host hidden moons. While \citet{bro13} argued that the lack of a satellite for Makemake suggested that it had escaped a past giant impact, the discovery of S/2015 (136472) 1 suggests that unless it resulted from the capture of a previously-unbound TNO, it too was subjected to a giant impact and its density will likely reflect that \citep{ste12}. The apparent ubiquity of trans-Neptunian dwarf planet satellites further supports the idea that giant collisions are a near-universal fixture in the histories of these distant worlds.

\section{Acknowledgements}

The authors would like to thank Simon Porter, John Spencer, and Leslie Young for helpful discussion on the tidal and thermal properties of minor planets and for constructive comments on drafts of this manuscript.

\begin{deluxetable*}{cccccc}
\tablecaption{S/2015 (136472) 1 discovery astrometry and photometry. \label{tab1}}
\tablenum{1}
\tablehead{
\colhead{JD$_{mid}$} & \colhead{mag\tablenotemark{a}} & \colhead{$\Delta\lambda$\tablenotemark{b}} & \colhead{$\Delta\beta$\tablenotemark{c}} & \colhead{$\Delta$R.A.\tablenotemark{d}} & \colhead{$\Delta$Dec\tablenotemark{e}}
}
\startdata
2457140.07994 & 25.21 & -0\arcsec.136 & 0\arcsec.556 & 0\arcsec.125 & 0\arcsec.558\\
2457140.10045 & 25.06 & -0\arcsec.135 & 0\arcsec.560 & 0\arcsec.127 & 0\arcsec.561\\
2457140.14283 & 25.19 & -0\arcsec.138 & 0\arcsec.555 & 0\arcsec.123 & 0\arcsec.558\\
2457140.15309 & 25.27 & -0\arcsec.140 & 0\arcsec.553 & 0\arcsec.120 & 0\arcsec.558\\
2457140.16334 & 25.12 & -0\arcsec.136 & 0\arcsec.554 & 0\arcsec.124 & 0\arcsec.557\\
\enddata
\tablenotetext{a}{AB-mag in HST WFC3 UVIS F350LP filter. Mean Makemake F350LP magnitude in four 12 second exposures: 17.39.}
\tablenotetext{b}{Sky-plane offset in Ecliptic Longitude, secondary to primary; $\Delta\lambda=(\lambda_{2} - \lambda_{1})\times\cos(\beta_{1})$.}
\tablenotetext{c}{Sky-plane offset in Ecliptic Latitude, secondary to primary; $\Delta\beta=\beta_{2} - \beta_{1}$.}
\tablenotetext{d}{Sky-plane offset in J2000 R.A., secondary to primary; $\Delta$R.A.=(R.A.$_{2} - $R.A.$_{1}) \times \cos$(Dec$_{1}$).}
\tablenotetext{e}{Sky-plane offset in J2000 Dec, secondary to primary; $\Delta$Dec=Dec$_{2} - $Dec$_{1}$.}
\end{deluxetable*}

\end{document}